\documentstyle[aps,epsfig,multicol]{revtex}
\begin{document}
%\onecolumn
\title{Wave packet evolution approach to ionization of hydrogen molecular
ion by fast electrons}
\author{Vladislav V. Serov, Vladimir L. Derbov}
\address{Chair of Theoretical and
Nuclear Physics, Saratov State University, 83 Astrakhanskaya,
Saratov 410026, Russia}
\author{Boghos B. Joulakian}
\address{Institut de Physique, Laboratoire de
Physique Mol\'eculaire et des Collisions, Universit\'{e} de Metz,
Technop\^{o}le 2000, 1 Rue Arargo, 57078 Metz Cedex 3, France}
\author{Sergue I. Vinitsky}
\address{Laboratory of Theoretical Physics, Joint Institute for
Nuclear Research, Dubna 141980, Moscow Region, Russia}
\date{\today}
\maketitle
\begin{abstract}
 The multiply differential cross section of the
ionization of hydrogen molecular ion by fast electron impact is
calculated by a direct approach, which involves the reduction of
the initial 6D Schr\"{o}dinger equation to a 3D evolution problem
followed by the modeling of the wave packet dynamics. This
approach avoids the use of stationary Coulomb two-centre functions
of the continuous spectrum of the ejected electron which demands
cumbersome calculations.  The results obtained, after verification
of the procedure in the case atomic hydrogen, reveal interesting
mechanisms in the case of small scattering angles.\\[3mm] PACS
number(s): 34.80.Dp
\end{abstract}

\begin{multicols}{2}
\section{Introduction}
New experimental methods, particularly, based on the multiple
coincidence detection technique \cite{Wang89,Corchs93,Corchs94}
stimulate the interest to fundamental theoretical studies of the
dissociative ionization of diatomic molecules by electron impact.
In this context the molecular hydrogen ion can be considered as
the basic system in which the removal of the unique electron
causes dissociation. Substantial theoretical analysis of the
dissociative ionization of ${\rm H}^{+}_{2}$  by fast electrons
was recently carried out in \cite{Joulakian96}.  As mentioned in
\cite{Joulakian96}, the crucial point of calculating the
cross-section of such processes is that no closed exact
analytical wave functions of the continuum states exist. In
\cite{Joulakian96} the final-state wave function of the ejected
electron was found by taking a product of two approximate
functions that take into account the two scattering centers. To
improve the calculation it seems straightforward to obtain these
functions with  the  exact numerical solutions of the two-center
continuum problem. However, this approach involves a cumbersome
procedure of calculating multi-dimensional integrals of the
functions presented numerically that requires huge computer
facilities and may cause additional computational problems. It
seems reasonable to search for direct computational approaches,
in which the basis of exact two-center continuum wave functions
is not involved. Note that the potential advantage of such
methods is that they could be generalized over a  wider class of
two-center systems starting from the molecular hydrogen ion as a
test object. In the present paper we develop a direct approach to
the  ionization of hydrogen molecular ion by fast electrons that
involves the reduction of the initial 6D Schr\"odinger equation
to a 3D evolution problem followed by modeling of the wave packet
dynamics.

Originally we intended to treat the incoming electron
classically, its trajectory being approximated by a straight line
with the deflection neglected.  The bound electron was to be
treated  quantum mechanically. Preliminary calculations at the
impact parameter $\rho=10$ a.u. has shown, first, that the
probability of the emission of the electron having the energy of
50 eV is extremely small, and, second, that the direction of the
electron emission is orthogonal to that of the incoming electron
motion, that contradicts the results of \cite{Joulakian96}. This
means that the main contribution to the small-angle scattering
comes from the central collisions with the bound electron in the
region of its localization. Generally, the classically estimated
deflection angle of $1^{o}$ for scattered electron corresponds to
the impact parameter of the order  of 1 a.u., so that the
trajectory passes through the molecule and the classical
treatment of the incoming electron is not valid.

Here we develop and apply a direct approach to the calculation of
the angular distribution of scattered and ejected electrons that
involves the reduction of the initial 6D Schr\"odinger equation to
a 3D evolution problem followed by modeling of the wave packet
dynamics. The approach does not make use of the basis of
stationary Coulomb two-center functions of the continuous spectrum
for the ejected electron, whose proper choice is a crucial point
of other model calculations. Our approach can be considered as the
linearized version of the phase function method
\cite{Babikov68,Calogero67} for the multi-dimensional scattering
problem. The evolution problem is solved using the method based on
the split-step technique \cite{Marchuk} with complex scaling,
recently proposed by some of us and tested in paraxial optics
\cite{Serov99}. In the present paper the method as a whole is also
tested using the well known problem of electron scattering by
hydrogen atom \cite{Mott65}.

\newcommand{\nn}{\nonumber \\ }
\section{Basic equations}
We start from the 6D stationary Schr\"odinger equation  which
describes two electrons in the field of two fixed protons
\begin{equation}
 \left[H_0({\bf r})
-\frac{1}{2}\nabla^2_{\bf R}+V({\bf r},{\bf R}) \right]\Psi({\bf
r},{\bf R})=E \Psi({\bf r},{\bf R}), \label{Schr6d}
\end{equation}
\begin{figure}
\begin{center}
\includegraphics[width=0.45\textwidth]{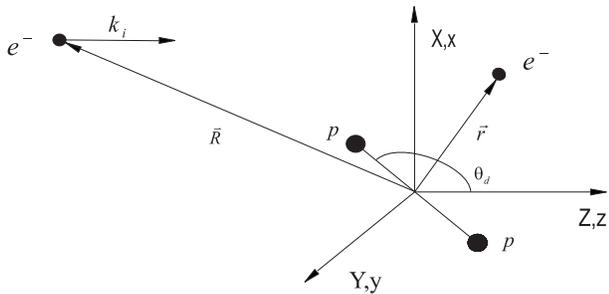}
\end{center}
\caption{Coordinate frame}
\end{figure}
where ${\bf r}$ is the radius-vector of the electron initially
bound in ${\rm H}_2^+$ and finally ejected, ${\bf R}$ is the
radius-vector of  the impact electron, $\hat
H_0=-\frac{1}{2}\nabla^2_{\bf r}+U({\bf r})$ is Hamiltonian of
ejected electron in the  field of two protons, $V({\bf r},{\bf
R})=U({\bf R})+U_{int}({\bf r},{\bf R})$ is the interaction
between the impact electron and molecular ion, $U({\bf
r})=-1/r_1-1/r_2$ is the attractive potential between the ejected
(scattered) electron and the protons,
 $r_1=|{\bf r}-{\bf r}_{1p}|$,  $r_2=|{\bf r}-{\bf r}_{2p}|$,
 ${\bf r}_{ip}$ is the radius-vector of the i-th proton,
 $U_{int}({\bf r},{\bf R})=1/|{\bf r}-{\bf R}|$ is the
repulsive potential of interaction between the electrons. The
origin of the coordinate frame is chosen in the center of symmetry
of the molecular ion with the $Z$ axis directed along the momentum
of the incident electron.
%$2{\bf d}={\bf r}_{2p}-{\bf r}_{1p}$,

For the scattering problem solved here the energy of the system
may be presented as $E=k_i^2/2+E_0$,  where $-E_0$ is the
ionization potential, $k_i$ is the momentum of the incident
electron.
 Let us seek  the solution of Eq.(\ref{Schr6d}) in the form
  $\Psi({\bf r},X,Y,Z)=\psi({\bf
r},{\bf R}_{\perp},Z)\exp(ik_{i}Z)$. Under the condition that
$(k_e^2+k_{\perp}^2-2E_0)/k_i^2<<1$ one can neglect the second
derivative of $\psi$ with respect to $Z$. As a result we get the
evolution-like equation for the envelope function $\psi({\bf
r},{\bf R}_{\perp},Z)$

\begin{eqnarray}
\lefteqn{ ik_i\frac{\partial \psi({\bf r},{\bf
R}_{\perp},Z)}{\partial Z}=} \nn & & \left\{ \hat H_0({\bf
r})-\frac{1}{2}\nabla^2_{{\bf R}_{\perp}}-E_0+V({\bf r},{\bf R})
\right\}\psi({\bf r},{\bf R}_{\perp},Z). \label{evolution}
\end{eqnarray}
Neglecting the large-angle scattering one can write the initial
condition for $\psi$ as
\begin{equation}
\psi({\bf r},{\bf R}_{\perp},-\infty)=\psi_0({\bf r}).
\end{equation}

To solve the 5D Schr\"odinger evolution equation(\ref{evolution})
we use Fourier transformation with respect to the variable
 ${\bf R}_{\perp}$
\begin{equation}
\psi({\bf r}, {\bf R}_{\perp}, Z)= \frac{1}{2\pi}\int\psi_{{\bf
k}_{\perp}}({\bf r},Z)\exp(i{\bf k}_{\perp}{\bf R}_{\perp}) d{\bf
R}_{\perp}.
\end{equation}

Then Eq.(\ref{evolution}) takes the form
\begin{eqnarray}
ik_i\frac{\partial \psi_{{\bf k}_{\perp}}({\bf r},Z)}{\partial Z}=
\left\{ \hat H_0({\bf r})+\left(\frac{k_{\perp}^2}{2}-E_0\right)
\right\}\psi_{{\bf k}_{\perp}}({\bf r},Z) && \nn
+\frac{1}{(2\pi)^2}\int V_{{\bf k}_{\perp}{\bf k}_{\perp}'}({\bf
r},Z) \psi_{{\bf k}_{\perp}'}({\bf r},Z)d{\bf k}_{\perp}', &&
\label{PrepBasicEq}
\end{eqnarray}
where
\begin{equation}
V_{{\bf k}_{\perp}{\bf k}_{\perp}'}({\bf r},Z)= \int\exp(-i({\bf
k}_{\perp}-{\bf k}_{\perp}'){\bf R}_{\perp}) V({\bf r},{\bf
R}_{\perp},Z)d{\bf R}_{\perp}
\end{equation}
is the Fourier transform of the interaction potential $V({\bf
r},{\bf R}_{\perp},Z)$.

Further simplification of the problem is possible if the
amplitude of the incident wave is much greater than that of the
scattered wave. In this case one can put
 \begin{equation}
\psi_{{\bf k}_{\perp}}({\bf r},Z)=  \delta({\bf
k}_{\perp})\psi_0({\bf r}) \label{born}
\end{equation}
 in the integral term of Eq.(\ref{PrepBasicEq}).
 As a result we get the inhomogeneous equation
\begin{eqnarray}
ik_i\frac{\partial \psi_{{\bf k}_{\perp}}({\bf r},Z)}{\partial
Z}&=& \left\{ \hat H_0({\bf
r})+\left(\frac{k_{\perp}^2}{2}-E_0\right) \right\}\psi_{{\bf
k}_{\perp}}({\bf r},Z)\nn &+&\frac{1}{(2\pi)^2}V_{{\bf
k}_{\perp}}({\bf r},Z)\psi_0({\bf r}), \label{BasicEq}
\end{eqnarray}
where $V_{{\bf k}_{\perp}}({\bf r},Z)=  V_{{\bf k}_{\perp}{\bf
0}}({\bf r},Z)$, with the initial condition $\psi_{{\bf
k}_{\perp}}({\bf r},-\infty)=0$.

To calculate the integral with respect to transverse variables in
the expression for
 $V_{{\bf
k}_{\perp}}({\bf r},Z)$ it is easier to start from the known
integral
\begin{equation}
\int\exp(-i{\bf k} {\bf R})\frac{1}{R}d{\bf R}=
\frac{4\pi}{k^2}=\frac{4\pi}{k_Z^2+k_{\perp}^2}.
\end{equation}
Carrying out the inverse Fourier transformation
\begin{equation}
\int_{-\infty}^{\infty}\exp(ik_Z
Z)\frac{dk_Z}{k_Z^2+k_{\perp}^2}=
\frac{\pi}{k_{\perp}}e^{-k_{\perp}|Z|},
\end{equation}
one gets
\begin{eqnarray}
\lefteqn{ V_{{\bf k}_{\perp}}({\bf r},Z)=\frac{ 2
\pi}{k_{\perp}}e^{-k_{\perp}|Z-z|-i{\bf k}_{\perp}{\bf
r}_{\perp}}}\nn &&-\frac{ 2 \pi}{k_{\perp}}\left[
 e^{-k_{\perp}|Z-d_Z|-i{\bf k}_{\perp}{\bf d}_{\perp}}
+e^{-k_{\perp}|Z+d_Z|+i{\bf k}_{\perp}{\bf d}_{\perp}}\right].
\label{MatrixEl}
\end{eqnarray}
Here $k_{\perp}=k_i\sin\theta_s$ is the transverse momentum
component of the scattered electron, $\theta_s$ is the scattering
angle, $\pm {\bf d}$ are the positions of the nuclei with respect
to the center of symmetry. Note that terms in square brackets
determine the elastic scattering of the incident electron by the
nuclei.

 Due to the exponential decrease of the source term
with $|Z|$ the integration may be actually carried out within a
certain finite interval $(-Z_{max},Z_{max})$. Hence the zero
initial condition should be imposed at the point $-Z_{max}$.

Note that the approximation (\ref{born}) is
 actually equivalent to the first Born approximation
 \cite{Mott65}. Multiply Eq.(\ref{BasicEq}) by the complex
 conjugate function of the continuous spectrum of $\hat H_0$ and
 integrate over all ${\bf r}$. Then
\begin{eqnarray}
\lefteqn{ik_i\frac{dC_{{\bf k}_{\perp}}({\bf
k}_e,Z)}{dZ}=\left\{\frac{k_e^2}{2}+\frac{k_{\perp}^2}{2}-E_0\right\}
C_{{\bf k}_{\perp}}({\bf k}_e,Z)} \nn &&
+\frac{1}{(2\pi)^2}\int\psi^*({\bf k}_e,{\bf r})V_{{\bf
k}_{\perp}}({\bf r},Z)\psi_0({\bf r})d{\bf r},
\end{eqnarray}
where  $C_{{\bf k}_{\perp}}({\bf k}_e,Z)=\int\psi^*({\bf k}_e,{\bf
r})\psi_{{\bf k}_{\perp}}({\bf r},Z)d{\bf r}$ is the probability
density amplitude for the transition of the initially bound
electron into the state with the momentum
 ${\bf k}_e$. Let us substitute $$C_{{\bf
k}_{\perp}}({\bf k}_e,Z)= \tilde{C}_{{\bf k}_{\perp}}({\bf
k}_e,Z)\exp(ik_Z Z),$$ where  $k_Z$ is the increment of the
longitudinal component of the momentum of the impact electron
determined by the relation
\begin{equation}
k_Z=-\frac{1}{k_i}\left(\frac{k_e^2}{2}+\frac{k_{\perp}^2}{2}-E_0\right).
\end{equation}
This relation is actually equivalent to the energy conservation
law written neglecting the terms of the order of   $k_Z^2$. The
substitution yields
\begin{eqnarray}
\lefteqn{ik_i\frac{d\tilde{C}_{{\bf k}_{\perp}}({\bf
k}_e,Z)}{dZ}=} \nn & & \frac{1}{(2\pi)^2}\exp(-ik_Z
Z)\int\psi^*({\bf k}_e,{\bf r})V_{{\bf k}_{\perp}}({\bf
r},Z)\psi_0({\bf r})d{\bf r},
\end{eqnarray}
and
\begin{eqnarray}
\lefteqn{\tilde{C}_{{\bf k}_{\perp}}({\bf k}_e,\infty)=} \nn & &
\frac{1}{ik_i(2\pi)^2}\langle  e^{i{\bf k}_s{\bf R}} \psi({\bf
k}_e,{\bf r})|V({\bf r},{\bf R})| e^{i{\bf k}_i{\bf R}}\psi_0({\bf
r})\rangle ,\label{C}
\end{eqnarray}
where ${\bf k}_s={\bf k}_i-{\bf K}$ is the momentum of the
scattered electron,  ${\bf K}=(-k_X,-k_Y,-k_Z)$ is the momentum
transfer.

Provided that the ejected electron has the momentum ${\bf k}_e$,
the  asymptotic form of the solution of Eq. (\ref{Schr6d}) for the
wave function of the scattered electron when $R\rightarrow \infty$
is
\begin{equation}
\Psi_{{\bf k}_e}^{as}({\bf R})=\exp(ik_{i}Z)
 +\frac{\exp(ik_sR)}{R}f_{{\bf k}_e}(\theta_s,\phi_s).
 \label{asboundary}
\end{equation}
The  scattering differential cross-section(DCS) can be then
expressed as
\begin{equation}
\sigma_{{\bf k}_e}(\theta_s,\phi_s)= \frac{k_e
k_s}{k_i}\left|f_{{\bf k}_e}(\theta_s,\phi_s)\right|^2,
\label{sigma}
\end{equation}

On the other hand, the asymptotic form of the wave function
resulting from the solution of Eq.(\ref{BasicEq})  under the
condition $Z\rightarrow \infty$ can be presented as
\begin{eqnarray}
\lefteqn{\Psi_{{\bf k}_e}^{as}({\bf R})=\exp(ik_{i}Z)+} \nn &&
\exp(ik_{i}Z)\int \tilde{C}_{{\bf k}_{\perp}}({\bf
k}_e,\infty)\exp(i{\bf k}_{\perp}{\bf R}_{\perp}+ik_Z Z)d{\bf
k}_{\perp}. \label{asboundary1}
\end{eqnarray}
Making use of the fact that the integrand has a stationary point
we finally get
\begin{eqnarray}
\lefteqn{\Psi_{{\bf k}_e}^{as}({\bf R})=e^{ik_{i}Z}+} \nn & &
\frac{1}{Z}e^{i\left(k_i-\frac{k_e^2/2-E_0}{k_i}\right)Z+
i\frac{k_i}{2Z}R_{\perp}^2}(-2\pi i k_i) \tilde{C}_{{\bf
k}_{\perp}^0}({\bf k}_e,\infty), \label{asboundary2}
\end{eqnarray}
where ${\bf k}_{\perp}^0=k_i\sin\theta_s(\cos\phi_s,\sin\phi_s)$,
$R_{\perp}=R\sin\theta_s$, $Z=R\cos\theta_s$. The expression
(\ref{asboundary2}) agrees with (\ref{asboundary}) within the
accuracy of the order of $\theta_s^2$ if we set
\begin{eqnarray}
\lefteqn{f_{{\bf k}_e}(\theta_s,\phi_s)=-2\pi i k_i
\tilde{C}_{{\bf k}_{\perp}^0}({\bf k}_e,\infty)} \nn & &=
-\frac{1}{2\pi}\langle e^{i{\bf k}_s{\bf R}} \psi({\bf k}_e,{\bf
r})|V({\bf r},{\bf R})| e^{i{\bf k}_i{\bf R}}\psi_0({\bf
r})\rangle.\label{f}
\end{eqnarray}
The latter expression is similar to the formula for $f_{{\bf
k}_e}(\theta_s,\phi_s)$ derived in \cite{Mott65} using the first
Born approximation.

\section{Calculation  of the angular distribution}
The asymptotic expression of the radial part of the wave function
corresponding to the continuous spectrum of $\hat H_0$ can be
written as
\begin{equation}
\psi^{as}_E(r,t)=\frac{1}{\sqrt{\upsilon(r)}\,\,r}
\exp(-iEt+i\int^{r}\upsilon(r')dr')
\end{equation}
where $t=Z/k_i$ is the evolution variable,
$\upsilon(r)=\sqrt{2(E-U^{as}(r))}$,
$E=\frac{k_e^2}{2}+\frac{k_{\perp}^2}{2}-E_0$, $U^{as}(r)=-Z'/r$,
$Z'=2$ is charge of two protons. In the asymptotic limit one can
take only the radial component of the momentum of the ejected
electron into account. Then, according to \cite{Selin99},  the
expression for calculating the amplitude $A(k,\theta,\phi)$ takes
the form
\begin{eqnarray}
\lefteqn{A_{{\bf k}_{\perp}}(k_e,\theta_e,\phi_e) =} \nn & &
\frac{1}{\sqrt{2\pi}}\left.\int_{t_0}^{t_1}dt' j(\psi_{{\bf
k}_{\perp}}(r,\theta,\phi,t'),\psi^{as}_E(r,t'))\right|_{r=r_{max}},
\label{Amplitudes}
\end{eqnarray}
where
\begin{equation}
j(\Psi,\Phi)=\frac{i}{2}\left\{
\Psi \,r^2\frac{\partial \Phi^*}{\partial r}-
\Phi^* \,r^2\frac{\partial \Psi}{\partial r}
\right\},
\end{equation}
is the  flux introduced in \cite{Selin99}, $t_0=-Z_{max}/k_i$
and $t_1>>Z_{max}/k_i$.
The approximate relation (\ref{Amplitudes}) becomes exact when
$t_1\rightarrow +\infty$ and simultaneously $r_{max}\rightarrow
+\infty$.

The amplitudes  defined  by (\ref{Amplitudes}) are related with
 the coefficients  introduced  in Eq.(\ref{C}) by
\begin{equation}
\left|A_{{\bf k}_{\perp}}(k_e,\theta_e,\phi_e)\right|^2=
k_e\left|\tilde{C}_{{\bf k}_{\perp}}({\bf k}_e,\infty)\right|^2
\label{AC}
\end{equation}
Using  (\ref{sigma}),(\ref{f}) and (\ref{AC}) we get the  final
expression for the differential cross-section
\begin{equation}
\sigma_{{\bf k}_e}(\theta_s,\phi_s)=(2\pi)^2 k_s k_i \left|A_{{\bf
k}_{\perp}}(k_e,\theta_e,\phi_e)\right|^2.
\end{equation}

In the region where $r>r_{max}$ we made use of the complex
scaling technique \cite{ComplSc} to suppress the non-physical
reflection from the grid boundary.

\section{Numerical scheme}
 The inhomogeneous  Schr\"odinger equation to be solved can be
 written as
\begin{equation}
i\frac{\partial \Psi({\bf r},t)}{\partial t} =\hat H_0({\bf
r})\Psi({\bf r},t)+F({\bf r},t),\label{s1}
\end{equation}
The  solution of Eq. (\ref{s1}) to within  the second-order terms
in $\Delta t$ can be expressed as
 the following sequence of equations:
\begin{eqnarray}
\Psi^0_{l}\phantom{t+\Delta t)}&=&\Psi(t)-\frac{i\Delta t}{2}
F({\bf r},t);\label{CrNc1}\\
 (1+i\frac{\Delta t}{2}\hat
H_0)\Psi^1_{l}&=&(1-i\frac{\Delta t}{2}\hat
H_0)\Psi^0_{l};\label{CrNc2}\\
 \Psi(t+\Delta t)&=&\Psi^1_{l}-\frac{i\Delta t}{2} F({\bf r},t+\Delta
 t).\label{CrNc3}
 \end{eqnarray}
The key step of the procedure is  Eq. ~(\ref{CrNc2}) which
defines nothing but Cranck-Nicholson scheme. To solve this
equation we make use of the partial coordinate splitting (PCS). A
finite-difference scheme is applied for the radial variable $r$
and the polar angle $\theta$.  Fast Fourier transform (FFT) is
used  for the azimuthal angle $\phi$.

In the spherical coordinate system, the $z$-axis of which is
directed along the symmetry axis of the molecule (and not along
the velocity of the impact electron) and after the substitution
$\Psi=\psi/r$ and the Fourier transformation of Eq.(\ref{CrNc2})
the  terms $\hat H_0\Psi^{0,1}_{l}$ entering this equation turn
into
\begin{eqnarray}
\lefteqn{\hat H_0(r,\eta,m)\psi(r,\eta,m)=}\nn & &
-\frac{1}{2}\left[ \frac{\partial^2 }{\partial r^2}+
\frac{1}{r^2}\left( \frac{\partial }{\partial \eta}
(1-\eta^2)\frac{\partial }{\partial \eta}
-\frac{m^2}{1-\eta^2}\right) \right]\psi(r,\eta,m)\nn & &
+U(r,\eta)\psi(r,\eta,m),
\end{eqnarray}
where $\eta=\cos\theta$, $m$ is the asimuthal quantum number.

 Finite-difference approximation  $ \frac{\partial^2 }{\partial
r^2}\simeq D^{r}_{i_2 i_1}$ and $\frac{\partial }{\partial \eta}
(1-\eta^2)\frac{\partial }{\partial \eta}
-\frac{m^2}{1-\eta^2}\simeq D^{\eta,m}_{j_2 j_1}$ of the
differential operators entering Eq.(\ref{CrNc2}) yields  $M$ sets
of linear equations, each set being of the order $L\times N$,
where $M,\quad L$ and $N$ are the numbers of grid points in
$\phi$, $\eta$ and $r$, respectively. Direct solution of  each set
of
 equations requires $NL^2$ operations at each step
in $t$. The FFT that should be performed twice, first, when
proceeding from (\ref{CrNc1}) to (\ref{CrNc2}),  and second, from
(\ref{CrNc2}) to (\ref{CrNc3}), requires $N L M \log_2 M$ extra
operations.

 To reduce the number of operation we propose a
double-cycle split-step scheme. In case when $\hat H_0$ can be
presented as a sum $\hat H_0=\hat H_1+\hat H_2$, this  scheme can
be formulated as follows
\begin {eqnarray}
  \psi _ {1} &=& \psi (t); \nonumber \\
 (I + i\frac{\Delta t}{4} \hat H _ {1}) \psi_{2} &=&
( I-i\frac{\Delta t}{4} \hat H _ {1}) \psi_{1}; \nonumber
\\
 (I + i\frac{\Delta t}{4} \hat H _ {2}) \psi_{3} &=&
( I-i\frac{\Delta t}{4} \hat H _ {2}) \psi_{2}; \nonumber
\\
 (I + i\frac{\Delta t}{4} \hat H _ {2}) \psi_{4} &=&
( I-i\frac{\Delta t}{4}\hat H _ {2}) \psi_{3}; \nonumber
\\
 (I + i\frac{\Delta t}{4} \hat H _ {1}) \psi_{5} &=&
( I-i\frac{\Delta t}{4} \hat H _ {1}) \psi_{4}; \nonumber
\\
  \psi (t+\Delta t) &=& \psi_{5}, \nonumber
\end {eqnarray}
which to within  the second-order terms in $\Delta t$ corresponds
to  the initial Cranck-Nicholson scheme, $I$ and $H_{1,2}$ is
square matrixes,
 $(I)_{i2i1j2j1}=\delta_{i2i1}\delta_{j2j1}$.

Now the problem is how to split the Hamiltonian $\hat H_0$ into
two parts. Formal separation of radial and angular parts leads to
difficulties associated with the singularity of the angular part.
Due to this singularity the scheme appears to be conditionally
stable with severe limitations imposed on the step $\Delta t$.
Practically this version of the splitting scheme is applicable
only if the grid in $r$ is rough enough.

To remove this limitation we propose a partial coordinate
splitting scheme. Its principal idea is that in the vicinity of
$r=0$ it is preferable not to split off the angular part at all.
To implement this idea we introduce the $r$-dependent weight
function $p(r)$ which is supposed to diminish in the vicinity of
$r=0$ and define the discrete approximation of the operators
$\hat H_{1,2}$ in the following way
\begin{eqnarray}
 \lefteqn{(\hat H_1^m)_{i_2 i_1 j_2 j_1}= -\frac{1}{2}{\hat
D^{r}}_{i_2 i_1}\delta_{j_2 j_1}+U^{as}(r_i)}\nn & & +p(r_{i_1})
\left[-\frac{1}{2}\frac{
 {\hat D^{\eta,m}}_{j_2 j_1}}{r_{i_1}^2}\delta_{i_2 i_1}
+U_{2}(r_{i_1},\eta_{j_1})\right];\\
 \lefteqn{(\hat H_2^m)_{i_2 i_1
j_2 j_1}=}\nn & &
 (1-p(r_{i_1}))\left[-\frac{1}{2}
 \frac{{\hat D^{\eta,m}}_{j_2 j_1}}{r_{i_1}^2}\delta_{i_2 i_1}
 +U_{2}(r_{i_1},\eta_{j_1})\right],
\end{eqnarray}
here $U_{as}(r)+U_{2}(r,\eta)=U(r,\eta)$. It is reasonable to
choose $p(r)$ as a cubic polynomial
\begin{eqnarray}
p(r)=\left\{
\begin{array}{ll}
2\left[\frac{r-r_a}{a_p}\right]^3-3\left[\frac{r-r_a}{a_p}\right]^2+1,&r_a<r<r_a+a_p;\\
1,&r\leq r_a;\\ 0,&r\geq r_a+a_p;
\end{array}
\right. & & \nonumber
\end{eqnarray}
where $r_a$ is the radius of the vicinity of $r=0$ where the
splitting is absent, $a_p$ is the width of the area of partial
splitting. Such a polynomial satisfies the condition of smooth
connection at the boundaries that separate the region of partial
splitting from the  regions  of full splitting, on one hand, and
of no splitting at all, on another hand.

\section{Numerical calculations and results}
The method was tested using the well-studied example of the impact
ionization of atomic hydrogen. We compared our results with those
given by the well-known expression obtained  in the first Born
approximation \cite{Landau}.    Good agreement was demonstrated in
the energy interval of interest $ E_e $ from 1 to 3 a.u., $E_e$
being the energy of the ejected electron.
\begin{figure}
\includegraphics[width=0.45\textwidth,clip]{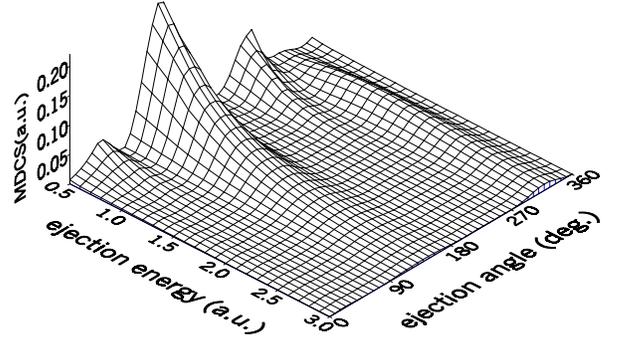}
\caption{The multi-fold differential cross section (MDCS) of the
ionization of $H_2^+$ versus the ejection angle $\theta_e$ and
ejection energy $E_e$ for $\theta_d=135^o$} \label{figqeEe}
\end{figure}
\end{multicols}
\begin{figure*}
\begin{center}
\includegraphics[width=0.45\textwidth]{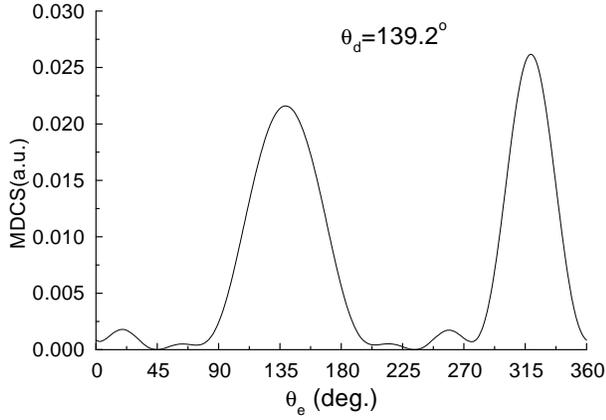}\hspace{1cm}
\includegraphics[width=0.45\textwidth]{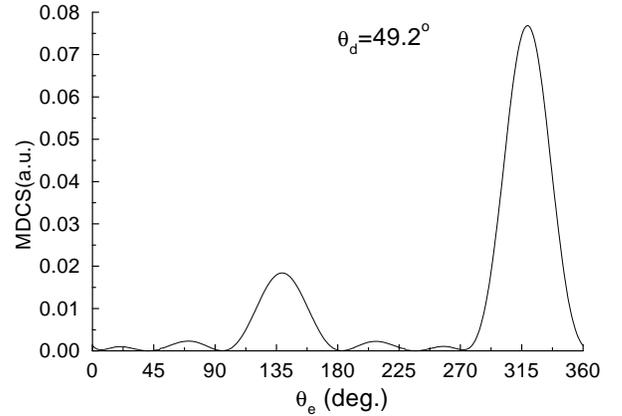}\\
\vspace{-3mm}
\parbox[b]{0.45\textwidth}{\center \hspace{-0.01\textwidth} \Large (a)}
\parbox[b]{0.45\textwidth}{\center \hspace{0.1\textwidth} \Large (b)}\\
\vspace{3mm}
\includegraphics[width=0.45\textwidth]{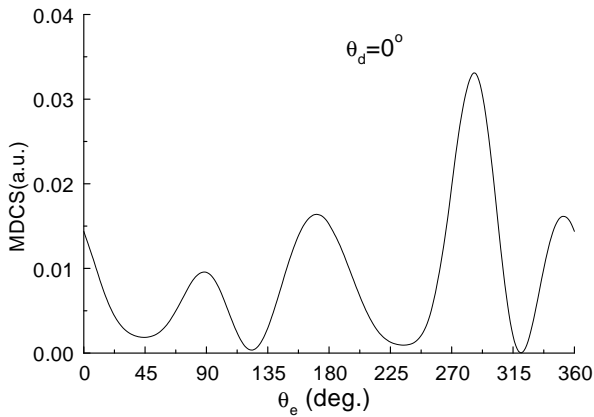}\hspace{1cm}
\includegraphics[width=0.45\textwidth]{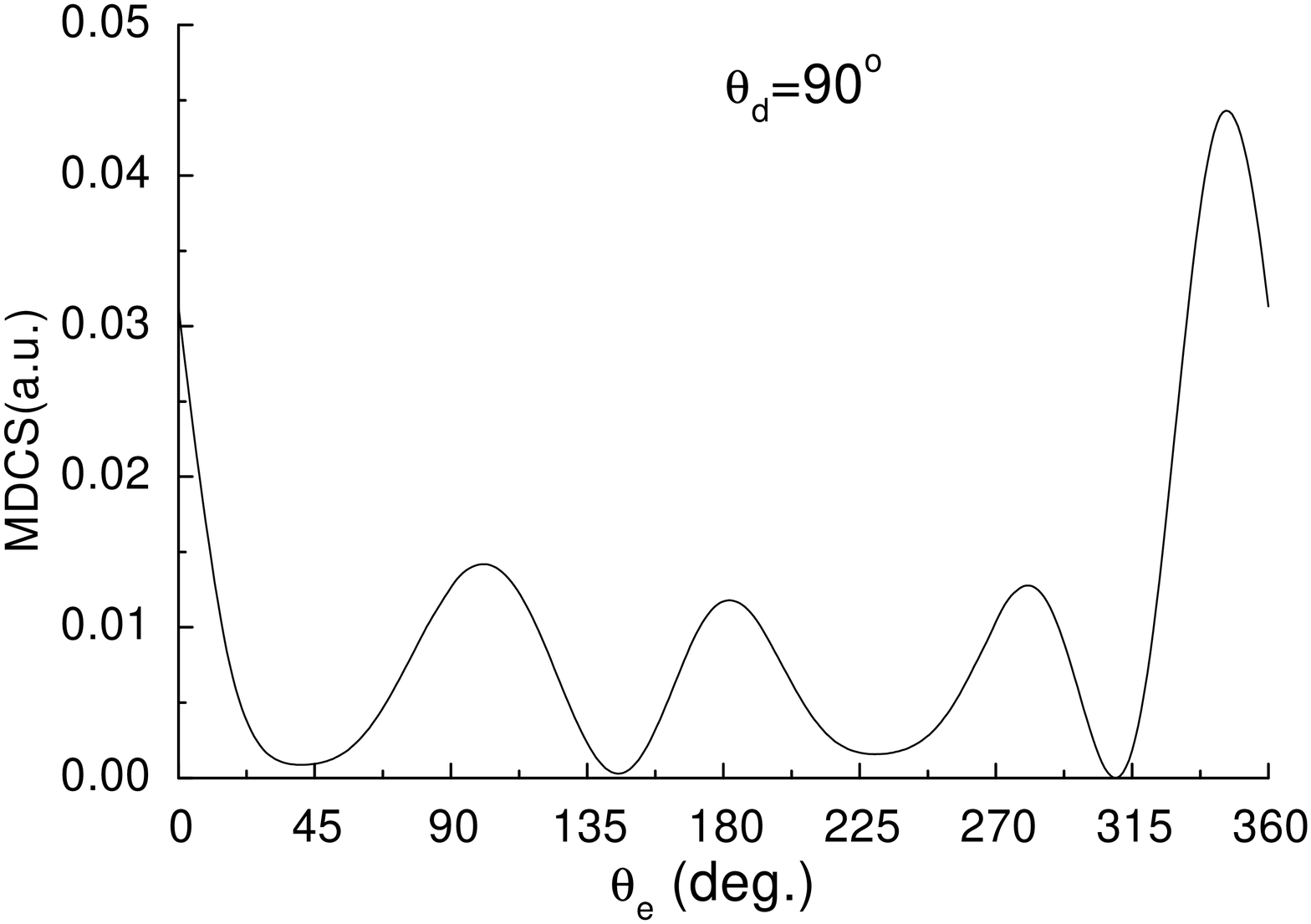}\\
\vspace{-3mm}
\parbox[b]{0.45\textwidth}{\center \hspace{-0.01\textwidth} \Large (c)}
\parbox[b]{0.45\textwidth}{\center \hspace{0.1\textwidth} \Large (d)}\\
\end{center}
\caption{The multi-fold differential cross section (MDCS) of the
ionization of $H_2^+$ versus the ejection angle $\theta_e$ for
different angles $\theta_d$\,: a)$\theta_d=139.2^o$ that
corresponds to ${\bf d}\parallel {\bf K}$\,; b)$\theta_d=49.2^o$
that corresponds to ${\bf d}\perp {\bf K}$\,; c)$\theta_d=0^o$; d)
$\theta_d=90^o$. The energy of the ejected electron $E_e=1.85$
a.u.=50.3 eV.} \label{fig4qe}
\end{figure*}
\begin{multicols}{2}
 Our numerical studies
concerning the molecular hydrogen ion focused on the variation of
the multi-fold differential cross section (MDCS) concerning a
coincidence detection of the two emerging electrons and one of the
protons with the ejection angle $\theta_e$ at different
orientations of the molecular axis, provided that the scattering
angle is small. The examples of our results illustrated by
Figs.\ref{figqeEe}-\ref{figqd} are obtained under the following
conditions: the momentum of the impact electron $k_i=$12.13 a.u.
($E_i\simeq 2000$ eV); the angle of scattering $\theta_s=1^o$. The
impact and ejected electron trajectories and the molecular axis
are supposed to lie in one plane.  The latter restriction is not
imposed by the method as such, it is just an example. Generally,
one gets full information about the ejected electron, i.e., the
dependence of MDCS from $E_e$, $\theta_e$ and $\phi_e$, after each
run of the code at given values of the impact energy, scattering
angle and molecular axis orientation. In Fig.\ref{figqeEe}
demonstrates the energy-angle distribution, extracted from the
data getting in result of one run of the code. In the planar
geometry the orientation of the molecular ion is determined by a
single angle $\theta_d$ between the impact direction and the
internuclear axis. We remind that the momentum transfer vector was
defined above as ${\bf K}={\bf k}_i-{\bf k}_s$. In Figs.
\ref{fig4qe} we present the particular cases of the dependence of
MDCS upon $\theta_e$ when internuclear axis is a)parallel to the
momentum transfer; b)perpendicular to the momentum transfer;
c)parallel to the impact electron direction ${\bf k}_i$;
d)perpendicular to the impact electron direction ${\bf k}_i$. As
it could be expected basing on the elementary symmetry
considerations,  the first two plots are symmetric with respect to
the direction of the momentum transfer that corresponds to the
angle $\theta_e=319.2^o$. Since this symmetry is not assumed {\it
a priori} in the procedure, this may be considered as one more
evidence in favour of the validity of the results demonstrated.
\begin{figure}
\includegraphics[width=0.45\textwidth,clip]{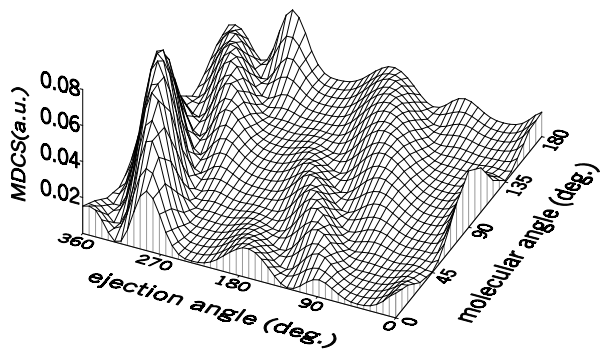}
\caption{The multi-fold differential cross section (MDCS) of the
ionization of $H_2^+$ versus the ejection angle $\theta_e$  and
molecular angle $\theta_d$. The energy of the ejected electron
$E_e=1.85$ a.u.=50.3 eV.} \label{figqeqd}
\end{figure}
The recoil momentum ${\bf Q}_{recoil}={\bf K} -{\bf k}_e$
transmitted to the target has its minimum for ${\bf k}_e$ parallel
to ${\bf K}$. In this case all the momentum is transferred to the
ejected electron and the probability of the ionization is maximal.
This is confirmed around $\theta_e =319.2^o$ on figures
\ref{fig4qe}(a) and \ref{fig4qe}(b) where the inter-nuclear axis
is respectively perpendicular and parallel to ${\bf K}$. So this
is a good verification for our calculation. Now, for the situation
where ${\bf k}_e$ is anti-parallel to ${\bf K}$, the recoil
momentum $Q_{recoil}$ is maximal and the probability of the
ionization is maximal. This is also visible for
$\theta_e=139.2^o$. Now for the directions of the internuclear
axis other than $\theta_d=139.2^o$ (where ${\bf d}$ is parallel to
${\bf K}$) or $\theta_d=49.2^o$ (where ${\bf d}$ is perpendicular
to ${\bf K}$) the target does not respect the above analysis. This
is due to the fact that the diatomic target behaves as an atomic
target only for these two angles. The other situations present
interference patterns the minima of which move when $\theta_d$
changes.

  Fig.\ref{figqeqd} shows MDCS  versus the
ejection angle $\theta_e$  and internuclear angle $\theta_d$. As
one can see, this dependence has rather a complex behaviour.
\begin{figure}
\includegraphics[width=0.45\textwidth,trim=0 -4mm 0 0,clip]{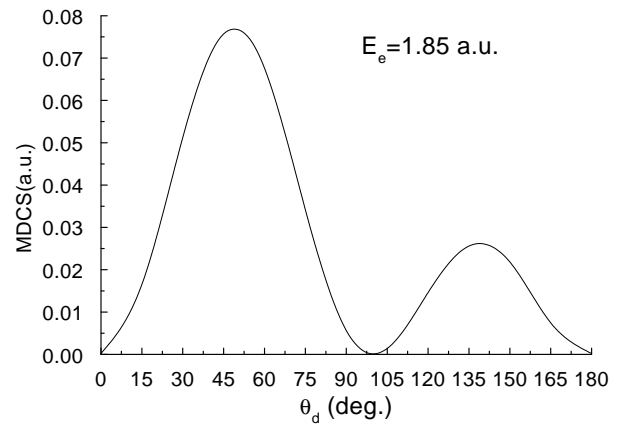}
\caption{The multi-fold differential cross section of the
ionization of $H_2^+$ as a function of the angle $\theta_d$
between the impact direction and the internuclear axis for fixed
ejection angle $\theta_e=319.2^o$. The energy of the ejected
electron is $E_e=1.85$ a.u.} \label{figqd}
\end{figure}
 To confirm the above dependence we show in Fig.\ref{figqd} a section
of Fig.\ref{figqeqd} for fixed ejection angle $\theta_e=319.2^o$
which corresponds to the case when the ejected electron direction
is parallel to the momentum transfer vector. It presents a
variation of the MDCS with respect to the direction of the
inter-nuclear axis. It can be clearly seen that the maximal value
of MDCS is achieved when the internuclear axis is perpendicular to
the momentum transfer direction that correspond to
$\theta_d=49.2^o$. This result agrees with the hypothesis
formulated in \cite{Bugacov}.
\section{Conclusion}
We have developed a procedure which determines the multiply
differential cross section of the (e,2e) ionization of hydrogen
molecular ion by fast electron impact, using a direct approach
which reduces the problem to a 3D evolution problem solved
numerically.  Our method avoids the cumbersome stationary
perturbative calculations, and opens the way for near future
applications to the (e,2e) ionization of more complex atomic and
molecular targets.

\acknowledgments Authors would like to thank Dr. A.V. Selin for
useful discussions. V.V.S and S.I.V. thanks to RFBR for supporting
by grants No-00-01-00617, No-00-02-16337.

\end{multicols}

\end{document}